\documentclass[5p,twocolumn,sort&compress,times,reversenotenum]{elsarticle}
\usepackage[colorlinks=true, pdfstartview=FitV, linkcolor=red,citecolor=blue, urlcolor=blue,
bookmarks=true, bookmarksnumbered=true, breaklinks]{hyperref}
\usepackage{color,graphicx}

\usepackage{amsmath,amssymb,tabularx,ulem,mathrsfs,bm,multirow,braket,fleqn}
\usepackage{fancyhdr}
\newcommand*{\e}[1]{
\begin{eqnarray}
#1
\end{eqnarray}
}

\newcommand{\n}[0]{
\nonumber\\}

\begin{document}

\begin{frontmatter}
\title{Hadronic Paschen-Back effect}
\author[1,2]{Sachio Iwasaki}\corref{cor1}
\ead{iwasaki.s.aa@m.titech.ac.jp}
\author[2]{Makoto Oka}
\ead{oka@post.j-parc.jp}
\author[3]{Kei Suzuki}
\ead{kei.suzuki@kek.jp}
\author[1]{Tetsuya Yoshida}
\ead{t.yoshida@th.phys.titech.ac.jp}

\address[1]{Department of Physics, Tokyo Institute of Technology, Meguro, Tokyo, 152-8551, Japan}
\address[2]{Advanced Science Research Center, Japan Atomic Energy Agency, Tokai, Ibaraki, 319-1195, Japann}
\address[3]{KEK Theory Center, Institute of Particle and Nuclear Studies, High Energy Accelerator Research Organization, 1-1, Oho, Ibaraki, 305-0801, Japan}

\cortext[cor1]{Corresponding author}

\begin{abstract}
We find a novel phenomenon induced by the interplay between a strong magnetic field and finite
orbital angular momenta in hadronic systems, which is analogous to the Paschen-Back effect observed
in the  field of atomic physics. This effect allows the wave functions to drastically deform. We discuss
anisotropic decay from the deformation as a possibility to measure the strength of the magnetic field in high-energy heavy-ion collisions
, which has not been measured experimentally.
As an example we investigate charmonia with a finite orbital angular momentum in a strong magnetic field.
We calculate the mass spectra and mixing ratios.
To obtain anisotropic wave functions, we apply the cylindrical Gaussian expansion method.
There we use different extention parameters for the parallel and transverse directions to the magnetic field.
\end{abstract}
\end{frontmatter}

\thispagestyle{fancy}
\rhead{KEK-TH-2031}
\cfoot{}
\renewcommand{\headrulewidth}{0.0pt}
\pagestyle{empty}

\section{Introduction}
The Paschen-Back effect (PBE) for a quantum system under a strong magnetic field is well-known in the field of atomic physics \cite{Paschen1921}.
The effect occurs when the strength of the magnetic field is larger than the scale of the spin-orbit (LS) coupling of the system, called the Paschen-Back (PB) region.
In the PB region, the state can be approximated by the state vector specified by $L_z$ and $S_z$, where $L_z$ and $S_z$ are the $z$ components (parallel to the magnetic field) of the orbital and spin angular momenta, respectively.

In this Letter, we consider {\it Hadronic  Paschen-Back effect} (HPBE) for hadronic systems composed of constituent quarks.
We note that hadrons are different from atomic systems since the quarks are confined in a confinement potential.
In this work, we consider the charmonia, the bound states of a charm quark and a charm antiquark.
Therefore we use the non-relativistic constituent quark model.
The HPBE should be seen in all the hadronic systems with a finite orbital angular momentum ($L \neq 0$).

Heavy-ion collision (HIC) experiments provide us with a chance to search quark/hadronic degrees of freedom under extreme environments such as high temperature, density and vorticity.
In particular, it is theoretically predicted that the strongest magnetic field in the present universe can be created at the Relativistic Heavy Ion Collider, RHIC ($|eB| \sim 0.1\ \mathrm{GeV}^2$ at most), and the Large Hadron Collider, LHC ($|eB| \sim 1.0 \ \mathrm{GeV}^2$ at most) \cite{Kharzeev:2007jp,Skokov:2009qp,Voronyuk:2011jd,Ou:2011fm,Bzdak:2011yy,Deng:2012pc,Bloczynski:2012en,Bloczynski:2013mca,Deng:2014uja,Huang:2015oca,Hattori:2016emy,Zhao:2017rpf}.
They are comparable to the typical scale of quantum chromodynamics (QCD), $\Lambda \sim 0.3 \ \mathrm{GeV}$.
On the other hand, there is no hard evidence of magnetic fields in HICs so far.
One of the reasons of difficulties in measuring a magnetic field would be its short lifetime.
In contrast, {\it relatively low-energy} collisions at the Super Proton Synchrotron (SPS) and the Beam Energy Scan (BES) program at RHIC can create a magnetic field with a long lifetime ($t \sim 2 \ \mathrm{fm}/c$) and a maximum strength of $|eB| \sim 0.01 \ \mathrm{GeV}^2$ (e.g. see Ref.~\cite{Skokov:2009qp} for SPS energy).
The HPBE suggested in this Letter will provide a prospective probe of not only strong magnetic fields at RHIC and LHC but also even {\it relatively weak} magnetic fields at SPS and RHIC-BES.
In particular, charmonia are quickly produced by nucleon-nucleon scattering in the initial stages of HICs, so that it can be a suitable probe of the magnetic fields.

\section{Formulation of HPBE} 
Before numerical simulations, we formulate HPBE for the P-wave charmonia.
In vacuum, the P-wave charmonia are classified by spin-singlet $h_c$ ($^1 \! P_1$) and spin-triplets $\chi_{c0}$ ($^3 \! P_0$), $\chi_{c1}$ ($^3 \! P_1$), and  $\chi_{c2}$ ($^3 \! P_2$), where the total angular momentum $J=L+S$, orbital angular momentum $L$, and spin angular momentum $S$ for $^{2S+1} \! P_J$ states are the good quantum numbers because of the spherical symmetry of the vacuum (note that $L_z$ and $S_z$ are not conserved due to the LS and tensor coupling).

With a magnetic field along the $z$ direction, spherical symmetry is broken and only $J_z$ is strictly conserved. When the magnetic field is stronger than the spin-orbit splitting,  {\it i.e.,} the PB region, 
$L_z$ and $S_z$ are also conserved approximately
\footnote{Precisely speaking, the existence of the tensor coupling mixes $L_z$ and $S_z$ even in the PB limit.}.
The eigenstates of the P-wave charmonia can be represented by the {\it PB configuration} as follows \footnote{As an alternative notation, we can also use $\ket{LL_z;SS_{z}} \equiv Y_{L L_z} \chi_{S S_z}$.}: 
\e{
\Psi_{L_z;S_{1z}S_{2z}}(\rho,z,\phi) = \Phi_{L_z}(\rho, z) Y_{1 L_z}(\theta,\phi)\chi(S_{1z}, S_{2z}),
\hspace{0.5cm}
\label{eq.PBconf}
}
where $\tan\theta=\rho/z$, $S_{1z} (S_{2z})$ is the third component of the spin of the charm (anticharm) quark, 
$Y_{1L_z}(\theta,\phi)$ is the spherical harmonics, 
$\chi(S_{1z},S_{2z})$ is the spin wave function, and $\Phi(\rho,z)$ is the spatial wave function in the cylindrical coordinate
\footnote{
Note that the configuration contains the $L=1,3,5,\cdots$ components.
Nevertheless, we can factorize the wave function as Eq.~(\ref{eq.PBconf}) because these partial waves with the same $L_z$ always have the factor of $e^{\pm i\phi}\sin\theta\propto Y_{1\pm1}$ or $\cos\theta\propto Y_{10}$.
}.
Since the spatial distributions of $L_z=\pm1$ and $0$ are different because of the factor of the spherical harmonics, the transition from the $\ket{J;LS}$ states in a weak field to the $\ket{L_z;S_{1z}S_{2z}}$ ones in a PB region is associated with deformation of the wave functions.

We emphasize that this is qualitatively different from the deformation of the S-wave charmonia ($\eta_c$ and $J/\psi$) in a magnetic field.
In fact, the wave functions of the ground states are not so sensitive to magnetic fields, and the deformation requires a strong magnetic field, $|eB| \sim 1.5 \ \mathrm{GeV}^2$, compared with the scale of charm-quark mass \cite{Suzuki:2016kcs,Yoshida:2016xgm}.
This is because the deformation of S-wave comes from only Landau levels (LLs) of charm quarks, while that of the P-wave is induced by HPBE as well as LLs (also see Table~\ref{Tab_origin}).
Therefore, the P-wave charmonia can be more sensitive to magnetic fields than the S-wave charmonia.

\begin{table}[t!]
\centering
\caption{Summary of the origin of wave-function deformation for S-wave and P-wave charmonia in a magnetic field and the relevant energy scales.}
\begin{tabular}{l|c|c}
\hline\hline
States & Origin & Relevant scale \\
\hline
S-wave & Quark LLs & $ \sqrt{eB} \sim  m_c$ \\
P-wave ($J_z=\pm2$) & Quark LLs & $ \sqrt{eB} \sim m_c$ \\
P-wave ($J_z=\pm1$) & HPBE & $ \sqrt{eB} \sim \braket{ V_{LS}}$ \\
       & Quark LLs & $ \sqrt{eB} \sim m_c$ \\
P-wave ($J_z=0$) & HPBE & $ \sqrt{eB} \sim \braket{ V_{LS}}$ \\
       & Quark LLs & $ \sqrt{eB} \sim m_c$ \\
\hline\hline
\end{tabular}
\label{Tab_origin}
\end{table}

\section{Numerical setup}
We utilize the constituent quark model in a magnetic field \cite{Alford:2013jva,Bonati:2015dka,Suzuki:2016kcs,Yoshida:2016xgm}.
The properties of the S-wave charmonia ($J/\psi$ and $\eta_c$) in a magnetic field are well understood from the numerical approaches in this model \cite{Alford:2013jva,Bonati:2015dka,Suzuki:2016kcs,Yoshida:2016xgm}.
Some of their properties were confirmed also by the analyses in an effective Lagrangian \cite{Cho:2014exa,Cho:2014loa,Yoshida:2016xgm} and QCD sum rules \cite{Cho:2014exa,Cho:2014loa}.
We start from
\e{
\hspace{-0.05cm}
 H
=
\sum_{i=1}^2 \left[
        \frac1{2m_c} \left(
                 \mbox{\boldmath $ p$}_i - q_i \mbox{\boldmath $A$}(\mbox{\boldmath $r$}_i)
        \right)^2
        - \mbox{\boldmath $ \mu$}_i \cdot \mbox{\boldmath $B$}
        +m_c
\right] + V(r),
\hspace{0.5cm}
}
where $m_c$ is the constituent quark mass, and $q_i$, $\mbox{\boldmath $ p$}_i$, $\mbox{\boldmath $ \mu$}_i=gq_i \mbox{\boldmath $ S$}_i/2m_c$ and $\mbox{\boldmath $ S$}_i$ are the electric charge, momentum, magnetic moment and the spin operator of the $i$ th charm quark, respectively.
Now we assume a uniform constant magnetic field, and then we choose the gauge: $\mbox{\boldmath $A$}(\mbox{\boldmath $r$}) = \frac12 \mbox{\boldmath $B$} \times \mbox{\boldmath $r$}$.
We rewrite the Hamiltonian above in terms of the center of mass and relative coordinates, $\mbox{\boldmath $R$} = (m_c\mbox{\boldmath $r$}_1+m_c\mbox{\boldmath $r$}_2)/M$ and $\mbox{\boldmath $r$}= \mbox{\boldmath $r$}_1 - \mbox{\boldmath $r$}_2$, where $M= 2m_c$ is the total mass of the two constituent charm quarks.
Here we just offset the coordinate so that the center-of-mass of the charmonium is at rest at $\mbox{\boldmath $R$}=\mbox{\boldmath $0$}$.
Hence we can factorize the total wave function into a component including only $\mbox{\boldmath $r$}$.
The relative Hamiltonian can be written as
\e{
{ H}_{\mathrm{rel}}
&=&
{ H}_{\mathrm{diag}} +  H_{\mathrm{m}}
+  V_{\mathrm{LS}} +  V_\mathrm{T}, 
\label{eq.hamil}
\\
{ H}_{\mathrm{diag}} &=&
        \left[  -\frac1{2\mu} \nabla^2 + \frac{q^2B^2}{8\mu}\rho^2 \right]
        + \sigma r - \frac 43 \frac{\alpha_s}r 
        \n
        &&\hspace{0.5cm}	+\frac{32\pi \alpha_s}{9m_c^2}\left( 
		\frac\Lambda{\sqrt{\pi}}\right)^3 
	\left( {\mbox{\boldmath $S$}}_1\cdot {\mbox{\boldmath $S$}}_2 \right) e^{-\Lambda^2r^2},
\label{eq.hamildiag}
\\
 H_{\mathrm{m}}
&=&
-\sum_{i=1}^2 \left( \mbox{\boldmath $\mu$}_i \cdot \mbox{\boldmath $B$}\right),
\label{eq.magneticmoment}
\\
 V_{\mathrm{LS}}
&=&
\frac1{m_c^2}  \left(
        2\alpha_s A_{\mathrm{LS}} \frac{1-e^{-\Lambda_{\mathrm{LS}}^2 r^2}}{r^3} - \frac \sigma{2r}
\right) {\mbox{\boldmath $L$}}\cdot {\mbox{\boldmath $S$} },
\\
 V_\mathrm{T}
&=&
\frac{4\alpha_s A_{\mathrm T}}{m_c^2}
\frac{1-e^{-\Lambda_{\mathrm T}^2r^2}}{3r^3} 
\left[ 3({\mbox{\boldmath $S$}}_1\cdot {\mbox{\boldmath $\hat r$}})({\mbox{\boldmath $S$}}_2\cdot  \hat{\mbox{\boldmath $r$}})- {\mbox{\boldmath $S$}}_1\cdot {\mbox{\boldmath $S$}}_2 \right],
\label{eq.Hrel}
}
where 
$\hat{\mbox{\boldmath $r$}} = {\mbox{\boldmath $r$}}/|{\mbox{\boldmath $r$}}|$.
Here we write the Hamiltonian using the cylindrical coordinate  $(\rho, z,\phi)$.
$\mu = m_c/2$ is the reduced mass.
The magnetic field is assumed to be parallel to the $z$-axis: $ \mbox{\boldmath $B$}  = (0,0,B)$.
$q=|q_c|= \frac23e$ is the electric charge of the charm quark.
$ \mbox{\boldmath $ S$}  =  \mbox{\boldmath $ S$}_1+ \mbox{\boldmath $ S$}_2$ is the total spin operator for the charmonium.
$\mbox{\boldmath $ L$}=\mbox{\boldmath $r$}\times\mbox{\boldmath $ p$}$ 
is the orbital angular momentum operator of relative motion between charm and anti-charm quark.
Note that the $\mbox{\boldmath $ L$}\cdot\mbox{\boldmath $B$}$ term cancels for the quarkonia bacause the masses of the two particles are the same and the total electric charge is zero.
As a result, the $J_z=\pm2$ components of quarkonia are degenerate, and $J_z=\pm1$ components also are.
Here, we emphasize that PBE will occur even though the coupling between orbital angular momentum and magnetic field vanishes.
We adopt the model parameters from Barnes {\it et al}. \cite{Barnes:2005pb}: $(\sigma, \alpha_s, \Lambda, m_c)=(0.1425 \mathrm{GeV}^2, 0.5461, 1.0946 \mathrm{GeV}, 1.4794 \mathrm{GeV})$.
To stabilize numerical computations, we smear the LS and the tensor terms by introducing smearing parameters $(\Lambda_\mathrm{LS},A_\mathrm{LS},\Lambda_\mathrm{T},A_\mathrm{T}) = (0.2\mathrm{GeV},7.3, 1.2\mathrm{GeV},1.2)$,
fixed so as to reproduce the experimental values of the masses of the 1P charmonia.

We do not consider the $B$-dependence of the potentials.
In fact, the anisotropy of the confinement potential in a magnetic field is indicated by phenomenological models \cite{Miransky:2002rp,Andreichikov:2012xe,Chernodub:2014uua,Rougemont:2014efa,Simonov:2015yka,Dudal:2016joz,Hasan:2017fmf,Singh:2017nfa} as well as lattice QCD simulations at zero \cite{Bonati:2014ksa} and finite temperature \cite{Bonati:2016kxj,Bonati:2017uvz}.
Implementation of such anisotropy on the potential model as Ref.~\cite{Bonati:2015dka} would be interesting, but in this work we focus on HPBE in weak magnetic fields where the deformation of the potential can be safely neglected.

The Hamiltonian (\ref{eq.hamil}) is numerically solved by the cyrindrical gaussian expansion method (CGEM) \cite{Suzuki:2016kcs,Yoshida:2016xgm}.

\section{Numerical results}
\begin{figure}[t!]
    \centering
    \includegraphics[clip, width=1.0\columnwidth]{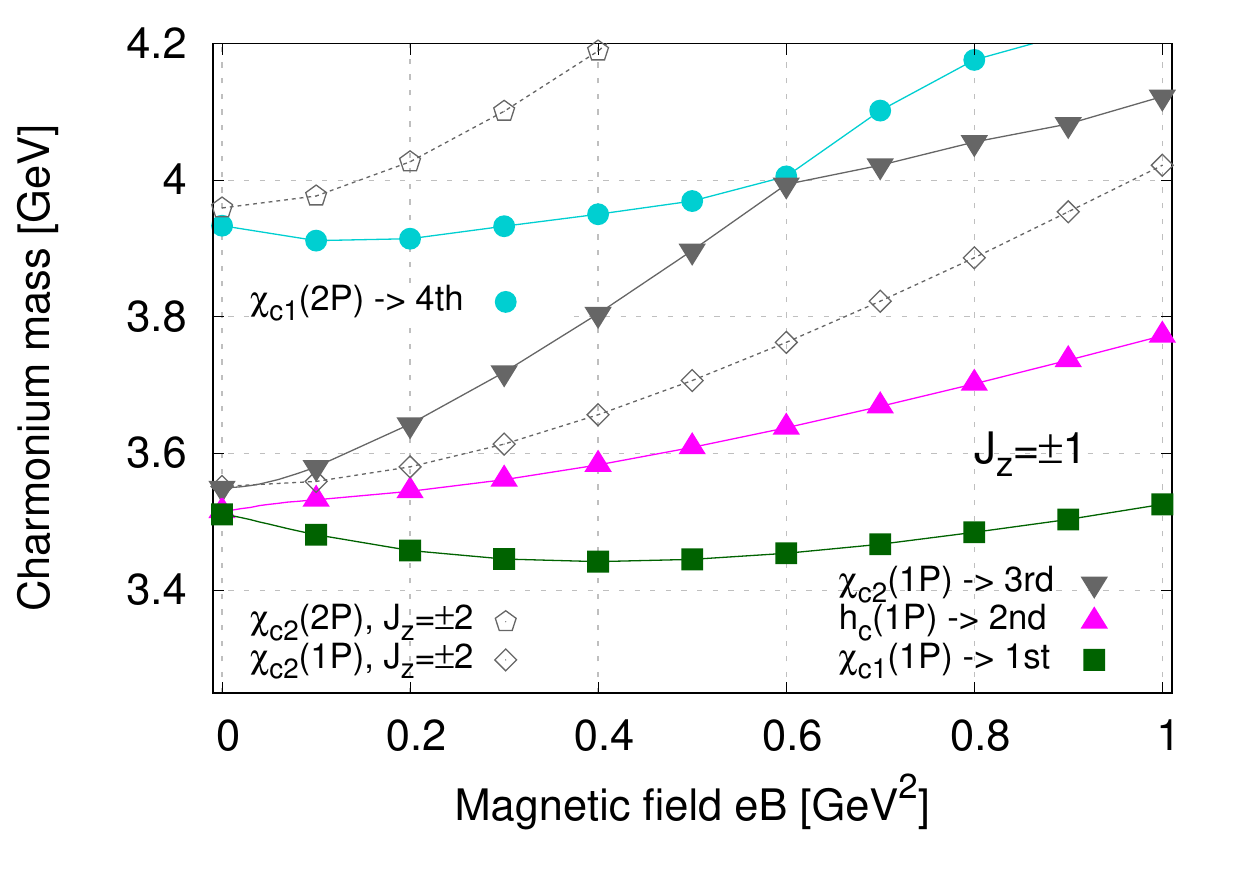}
    \includegraphics[clip, width=1.0\columnwidth]{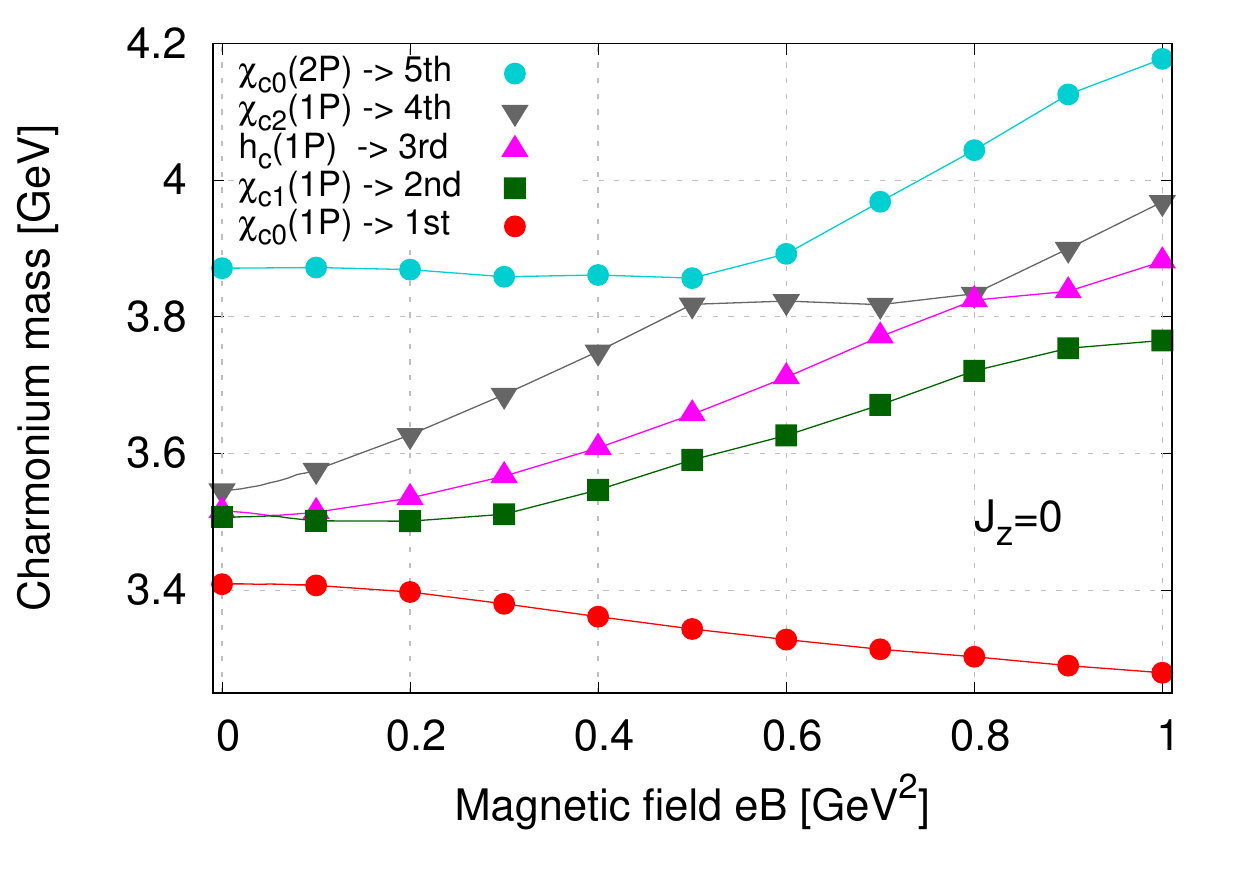}
    \caption{Masses of P-wave charmonia in a magnetic field. Upper: $J_z=\pm1$ and $\pm2$. Lower: $J_z=0$.}
    \label{Fig_mass}
\end{figure}

\begin{figure}[t!]
    \centering
    \includegraphics[clip, width=1.0\columnwidth]{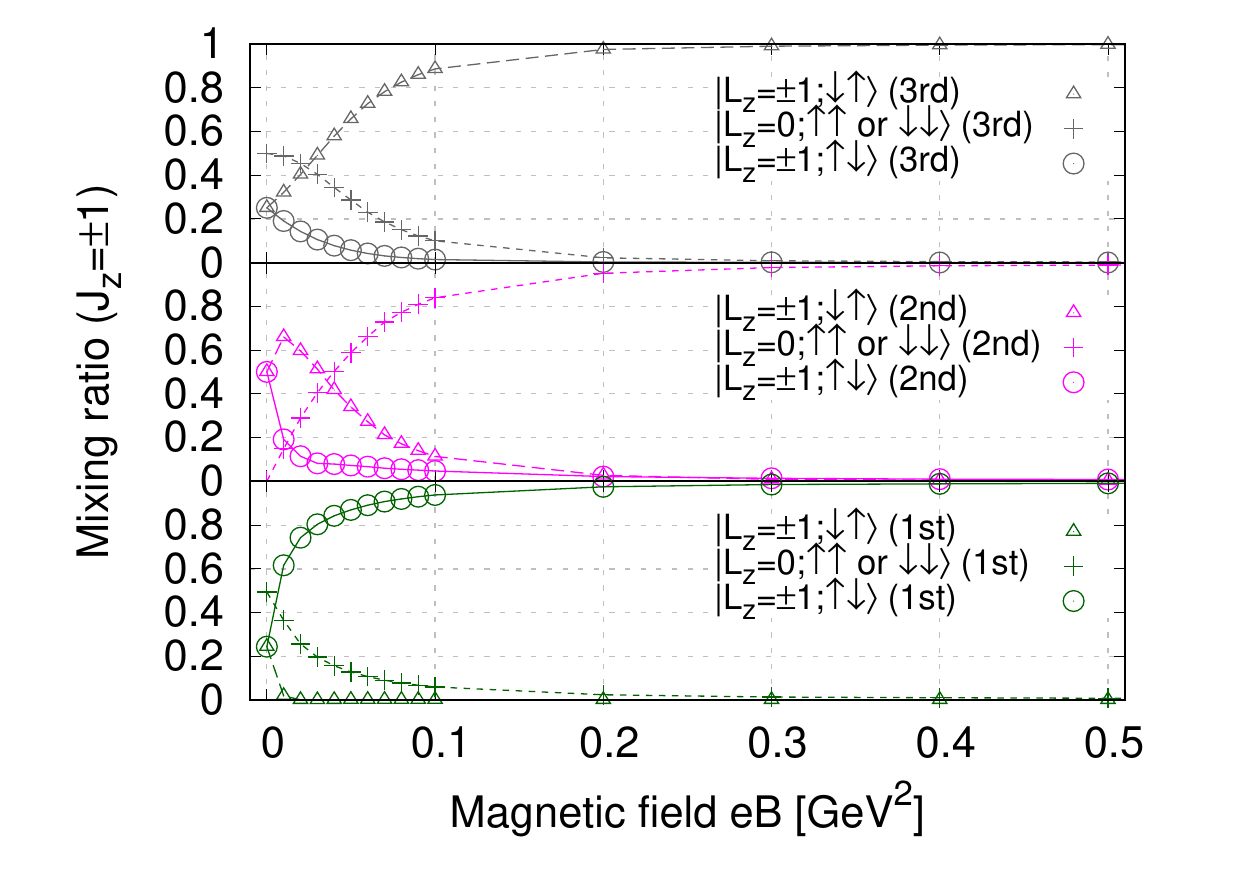}
    \includegraphics[clip, width=1.0\columnwidth]{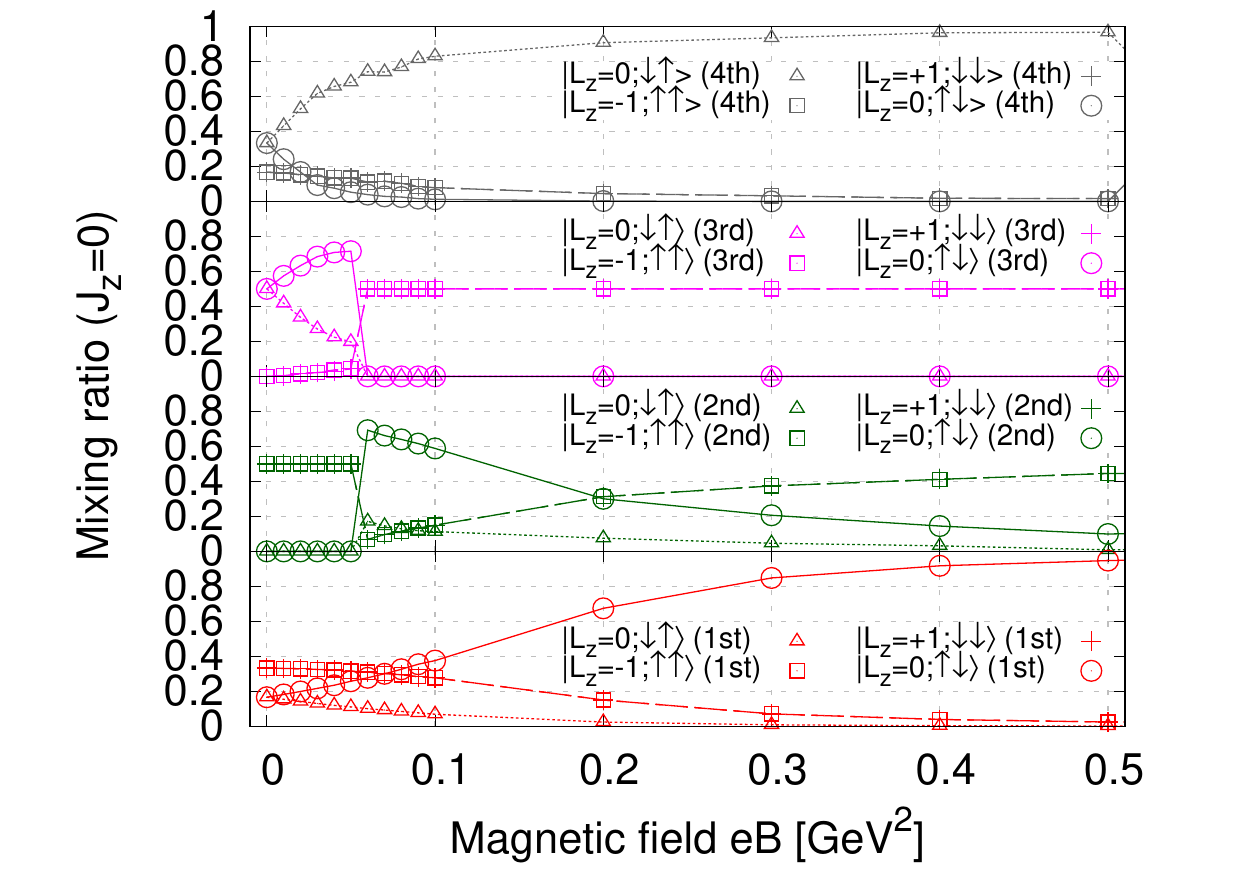}
    \caption{Mixing ratios among the $\ket{L_z;S_{1z}S_{2z}}$ basis of the P-wave charmonia in a magnetic field. Upper: $J_z=\pm1$. Lower: $J_z=0$.}
    \label{Fig_rate}
\end{figure}

The mass spectra are shown in Fig.~\ref{Fig_mass} and their mixing ratios in Fig.~\ref{Fig_rate}.
$ H_{\mathrm{m}}$ in Eq.~(\ref{eq.magneticmoment}) induces mixings between $S=0$ and $S=1$ states, and their levels repel each other.
Therefore, as the magnetic field gets stronger, the 1st state for $J_z=\pm1,0$ tends to fall, while
the 4th state in $J_z=0$ and 3rd state in $J_z=\pm1$, namely the heaviest states of 1P in each channel, goes upward rapidly.
The 1P state that most rapidly goes upward meets with the 2P state that goes downward, so that their levels cross.
We can see such level crossings  at $|eB| = 0.6\ \mathrm{GeV}^2$ in $J_z=\pm1$ channel and, $|eB| = 0.5$ and $0.8\ \mathrm{GeV}^2$ in $J_z=0$ channel.

Before we see the detail of the mass spectra, we move on to the mixing ratios in Fig.~\ref{Fig_rate} to confirm HPBE.
One sees that the saturation by the PB configurations is reached
already at the magnetic field around $|eB|\ge 0.2$ GeV$^2$.
We will explain the transition of mixing ratios in $J_z=\pm1$, and $0$.
Note that the $J_z=\pm2$ channels do not have mixing.

In vacuum, we can see the proper mixing ratios of the $\ket{L_z;S_{1z} S_{2z}}$ basis for each state.
For $J_z=\pm1$, the 1st, 2nd, and 3rd states correspond to $\chi_{c1}$, $h_c$, and $\chi_{c2}$, respectively.
For $J_z=0$, the four states are $\chi_{c0}$, $\chi_{c1}$, $h_c$, and $\chi_{c2}$, respectively.
The probabilities of $|L_z; S_{1z}S_{2z}\rangle$ states are given according to the Clebsch-Gordan
coefficients for the total $J$. 
When we turn on the magnetic field, one sees changes of the mixing ratios.
As the magnetic field gets stronger, the mixing between $S_z=0$ states, $\frac1{\sqrt2} (\ket{\uparrow \downarrow} - \ket{\downarrow \uparrow})$ and $\frac1{\sqrt2} (\ket{\uparrow \downarrow} + \ket{\downarrow \uparrow})$, increases.
As a result, the ratio of the $\ket{\uparrow \downarrow}$ ($\ket{\downarrow \uparrow}$) component in the lowest (highest) state of the 1P series converges into $1$ as shown in Fig.~\ref{Fig_rate}.
On the other hand, the other $S_z=\pm1$ states ($\ket{\uparrow \uparrow}$ and $\ket{\downarrow \downarrow}$) are left in the middle of 1P series, the 2nd state for $J_z=\pm1$, and the 2nd and 3rd for $J_z=0$.
Additionally, near $|eB|=0.05\ \mathrm{GeV}^2$, we can see that the mixing ratios switch because of the level crossing between the 2nd and 3rd in $J_z=0$.

\begin{figure}[t!]
    \centering
    \includegraphics[clip, width=1.0\columnwidth]{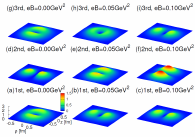}
    \caption{Probability densities of wave functions of P-wave charmonia for $J_z=\pm1$ in magnetic fields.
The vertical axis is $|\Psi_n(\rho, z, \phi)|^2$, and the horizontal plane is represented by the $\rho$ and $z$ axes, where $\rho$ ($z$) is the spatial direction perpendicular (parallel) to the magnetic field.
    }
    \label{Fig_WF}
\end{figure}

In Fig.~\ref{Fig_WF}, we show several examples of the density plots for $J_z=\pm1$.
At finite magnetic fields, we can see the clear spatial deformations.
At $|eB|=0.1\mathrm{GeV}^2$, we can see almost pure $|L_z|$ components.
In the case of $L_z=\pm1$, the basis functions have the factor of $r Y_{1\pm1}(\theta, \phi)\propto r\cos\theta e^{\pm i\phi}=\rho e^{\pm i\phi}$.
Then the wave functions become zero on the $z$-axis as the density plots in Fig.~\ref{Fig_WF}(c) and (i) show.
For $L_z=0$, the basis function is proportional to $r Y_{10}(\theta, \phi)\propto r\sin\theta=z$, and then the wave function is zero on the $\rho$-axis as shown in Fig.~\ref{Fig_WF}(f) .

Now we go back to the mass spectra.
In the mass shifts of the 1st states, we can see the characteristic behaviors:
the masses for $J_z=\pm1$ start to increase from $|eB| = 0.4\mathrm{GeV}^2$, and
that for $J_z=0$ keeps falling over $|eB| = 1.0\mathrm{GeV}^2$.
The difference comes from the spatial part of the wave functions.
The HPBE approximately fix $L_z$ in the 1st states:
$L_z=\pm1$ and $L_z=0$ in $J_z=\pm1$ and $J_z=0$, respectively.
Here the Hamiltonian in Eq.~(\ref{eq.hamildiag}) has the term of the harmonic oscillator potential in $\rho$ direction, which leads to the quark LLs.
Thus the wave functions are squeezed along $\rho$ direction as the magnetic field gets stronger.
Hence the $L_z=\pm1$ states, which are extended in $\rho$ direction, are squeezed and get higher energy, while the $L_z=0$ state stays at relatively lower energy, so that this level keeps going down by the level repulsion.

The $J_z=\pm2$ channels, the dotted lines in Fig.~\ref{Fig_mass}, have only the $L_z=\pm1$ component.
There is no mixing, and it shows the pure effect from the quark LLs.

Finally, we comment on the tensor coupling.
Without the tensor force, the 2nd and 3rd states in the $J_z=0$ spectrum are degenerate and become almost pure $L_z=\pm1$ states by the HPBE.
Since $\Delta L_z=2$ between them, they do not mix by the LS term, while they do by the tensor coupling.
The splitting by about $100$ MeV is given only by the tensor coupling.
Such a mass splitting would be important because it can provide a sensitive probe for the magnetic field.
For a large magnetic field, the wave function is squeezed and the matrix element of the tensor becomes larger because of the $1/r^3$ factor.
Thus this splitting will be sensitive to the magnetic field.
Also, the mixing ratio for the $\ket{\uparrow \uparrow}$ and $\ket{\downarrow \downarrow}$ components is always $0.5 : 0.5$ as shown in Fig.~\ref{Fig_rate}.

\section{Measurements of HPBE}
The HPBE can be related to the observables of the S-wave charmonia ~\cite{Marasinghe:2011bt,Tuchin:2011cg,Yang:2011cz,Tuchin:2013ie,Machado:2013rta,Dudal:2014jfa,Guo:2015nsa,Sadofyev:2015hxa,Suzuki:2016fof,Hoelck:2017dby,Braga:2018zlu}, heavy-light mesons \cite{Machado:2013rta,Machado:2013yaa,Gubler:2015qok,Yoshida:2016xgm,Reddy:2017pqp,Dhale:2018plh}, and heavy-quark diffusion \cite{Fukushima:2015wck,Finazzo:2016mhm,Das:2016cwd,Dudal:2018rki} in a magnetic field through the feed-down from the P-wave charmonia.
As a detectable effect of HPBE, we discuss the electric-dipole (E1) radiative decays of the P-wave charmonia.
E1 transitions change the orbital angular momentum by $\Delta L = \pm 1$, while they conserve the spin angular momentum, $\Delta S=0$.
In vacuum, the possible E1 decay processes are $h_c \to \eta_c \gamma$ and $\chi_c \to J/\psi \gamma$.
On the other hand, we should consider the decay processes in the $| L_z S_z \rangle$ basis in the magnetic field.
Now we consider the E1 transition amplitude between $^{2S+1}P$ and $^{2S+1}S$. 
The amplitude is given by $\braket{^{2S+1}S|\mbox{\boldmath $r$}\cdot \mbox{\boldmath $\epsilon$}^\pm|^{2S+1}P}$, where the polarization vector is given as $\mbox{\boldmath $\epsilon$}^{\pm} = \frac1{\sqrt2}(\pm1,-i\cos\alpha,-i\sin\alpha)$, and $\alpha$ is the angle between the directions of the magnetic field parallel to the $z$-axis and the photon momentum \footnote{In the coordinate system with the $z$-axis along the photon momentum vector, the polarization vector is $\mbox{\boldmath $\epsilon$}^{\pm \prime} = \frac1{\sqrt2}(\pm 1,-i,0)$. We rotated this by an angle $\alpha$ around the fixed $x$-axis to get $\mbox{\boldmath $\epsilon$}^{\pm}$.
}.
Since the spatial part of the wave function with $L_z=0$ is proportional to $z$, then we factorize it as $z\cdot \Phi_P(\mbox{\boldmath $r$};L_z=0)$.
Denoting the spatial part of the wave function of the S-wave as $\Phi_S(\mbox{\boldmath $r$})$, where the $\Phi_{S,P}(\mbox{\boldmath $r$})$ are even functions on $z$, we get
\e{
&&
\braket{S|\mbox{\boldmath $r$}\cdot \mbox{\boldmath $\epsilon$}^\pm|P;L_z=0}\n
\hspace{-1.0cm} &=&
\int \mathrm dV\ \Phi^*_S(\mbox{\boldmath $r$})
\cdot \frac1{\sqrt2} (\pm x-iy\cos\alpha -iz\sin\alpha)
z\Phi_P(\mbox{\boldmath $r$};L_z=0) \n
\hspace{-1.0cm} &=& 
- \sqrt2i\pi \sin\alpha \int \rho \mathrm d \rho \int \mathrm dz  
\ z^2\Phi^*_S(\mbox{\boldmath $r$})
\Phi_P(\mbox{\boldmath $r$};L_z=0) .
}
Thus photons cannot be emitted along the magnetic field due to the factor $\sin\alpha$.
Also for the $L_z=\pm1$ states, the spatial part of the wave function is proportional to $\mp \rho e^{\pm i\phi}$, so that
$\braket{S|\mbox{\boldmath $r$}\cdot \mbox{\boldmath $\epsilon$}^\pm|P;L_z=\pm1}$ is proportional to $(\cos\alpha\pm1)$ for $L_z=+1$ and to $(\cos\alpha\mp1)$ for $L_z=-1$.
These amplitudes indicate that the direction of the photons emitted from the P-wave charmonia shows angular dependence.
Thus the states with different $L_z$ emit photons with different angular distributions.
Then the HPBE can be measured through such anisotropic radiative decays.

Here we focused on only radiative decays.
However, the typical time scale of radiative decays should be comparable to that for the electromagnetic interactions: $\tau \sim 100 \, \mathrm{fm}/c$.
This might be too slow to be observed because of the short lifetime of magnetic fields in heavy-ion collision experiments.
On the other hand, as other observables, the time scales for the strong decays of quarkonia or quarkonium productions from heavy quarks should be that for the strong interaction: $\tau \sim 1 \, \mathrm{fm}/c$.
Such processes could be possible and ``more rapid" observables for the HPBE of P-wave quarkonia in the initial stage of heavy-ion collisions.

\section{Discussion and conclusion}
In this work we have focused on HPBE in a simplified situation with only a static and homogeneous magnetic field.
To consider more realistic situations in heavy-ion collisions, we examine the influence of (i) finite temperature, (ii) time evolution of magnetic field, (iii) finite vorticity, on HPBE for the P-wave charmonia.

(i) After the collision at RHIC and LHC, if quark gluon plasma (QGP) is produced, and its temperature is higher than the melting temperature of a charmonium ($T_{\bar{c}c} < T$), then the charmonium dissociates by the thermal effects, which is so-called charmonium suppression \cite{Matsui:1986dk}.
In QGP at lower temperature ($T_c < T < T_{\bar{c}c}$, where $T_c$ is the critical temperature of QCD), the charmonium survives, but the confinement potential is modified by the Debye-screening.
Under such a situation, implementation of the potentials modified by both the temperature and the magnetic field as estimated in Refs.~\cite{Rougemont:2014efa,Bonati:2016kxj,Bonati:2017uvz,Hasan:2017fmf,Singh:2017nfa} would be important. 
When QGP is not produced after the collision ($T < T_c$), and the P-wave charmonia do not suffer from the thermal effects (or can be slightly affected by thermal hadronic matter), then we can measure almost pure HPBE.

(ii) Naively, the strength of the magnetic fields at RHIC and LHC rapidly decreases as the spectator nuclei go away (unless we take into account a lasting mechanism \cite{Tuchin:2013ie,McLerran:2013hla,Tuchin:2013apa,Gursoy:2014aka,Zakharov:2014dia,Tuchin:2015oka} by the electric conductivity of QGP).
However, relatively low-energy collisions at SPS and RHIC-BNS can produce a long-lived magnetic field ($t \sim 2 \mathrm{fm}/c$).
HPBE for the P-wave charmonia, which is sensitive to even $|eB| \sim 0.01 \ \mathrm{GeV}^2$, could be a probe of magnetic fields.

(iii) Vorticity of produced nuclear/quark matter could be important, as recently observed at RHIC \cite{STAR:2017ckg}.
However, the operator of vorticity is represented by $\mbox{\boldmath $J$} \cdot \mbox{\boldmath $\Omega$}$, where $\bf{\Omega}$ is the vorticity, and it cannot mix the spin eigenstates of hadrons.
This is because vorticity, unlike magnetic fields, cannot distinguish positive or negative electric charges of quarks.
Therefore, we conclude that the qualitative properties of HPBE (and the anisotropic decay) are not affected by vorticity. 

The HPBE will occur in all the (nonrelativistic) mesonic systems with finite orbital angular momentum, (e.g. $h_1$-$\sigma (f_0)$-$f_1$-$f_2$, $b_1$-$a_0$-$a_1$-$a_2$, $K_0^\ast$-$K_1$-$K_2$, and $D_0^\ast$-$D_1$-$D_2$).
In particular, bottomonium systems such as $h_b$, $\chi_{b0}$, $\chi_{b1}$, and $\chi_{b2}$ can be created by heavy-ion collisions, and HPBE for such states could be also interesting.
For bottomonia, the LS coupling is smaller than that of charmonia because it is suppressed by the factor of the bottom-quark mass $1/m_b^2$.
As a result, the mass splitting due to the LS coupling becomes smaller (e.g. $\Delta m_{{h_c} - \chi_{c1}} \sim 15 \ \mathrm{MeV}$, while $\Delta m_{{h_b} - \chi_{b1}} \sim 6.5 \ \mathrm{MeV}$), which is a favorable situation for HPBE. 
On the other hand, the magnetic moment of bottom quarks, ${\boldsymbol \mu}_b \equiv g q_b {\bf S}/2m_b$, is smaller by the larger quark mass $m_b$ and the smaller electric charge $|q_b|=(1/3)e$, so that the spin mixing becomes weaker than the case of charmonia.
Thus HPBE for the P-wave bottomonia will be determined by the competition between these effects.

In summary, HPBE suggested in this work will be a good probe of the QCD physics under {\it relatively small} magnetic fields of the order of $|eB| \sim 0.01 \ \mathrm{GeV}^2$ (and also larger magnetic fields), which can be realized in heavy-ion collisions and compact stars.

\section*{Acknowledgment}
This work is partially supported by the Grant-in-Aid for Scientific Research (Grants No.~25247036 and No.~17K14277) from the Japan Society for the Promotion of Science.
K. S. is supported by MEXT as ``Priority Issue on Post-K computer" (Elucidation of the Fundamental Laws and Evolution of the Universe) and JICFuS.

\appendix
\section{The basis of CGEM for the P-wave states}
The Schr\"{o}dinger equation in this work is numerically solved by the cyrindrical gaussian expansion method (CGEM) \cite{Suzuki:2016kcs,Yoshida:2016xgm}.
In Refs.~\cite{Suzuki:2016kcs,Yoshida:2016xgm}, the basis functions for S-wave two-body systems in a magnetic field were introduced.
The spatial part of the basis functions for P-wave two-body systems with a fixed $L_z$ are as follows:
\e{
\Psi_n(\rho, z, \phi; L_z) &=& N_n r Y_{1L_z} (\theta, \phi)e^{-\beta_n\rho^2} e^{-\gamma_nz^2},
}
where $N_n$ is the normalization constant of the $n$ th basis, and $Y_{1L_z}(\theta, \phi)$ is the spherical harmonics.
$\beta_n$ and $\gamma_n$ are the range (variational) parameters which are optimized as the energy eigenvalue is minimized by the variational method.
Note that, for the spin part, the $\mbox{\boldmath $S$}_1 \cdot \mbox{\boldmath $S$}_2$ term gives the factors of $-3/4$ and $1/4$ for the $S=0$ and $1$ eigenstates, respectively. 

\begin{table}[t!]
\centering
\caption{Wave functions of P-wave charmonia {\it in vacuum}, represented by $Y_{LL_z} \chi_{S S_z}$ basis. }
\begin{tabular}{lcc}
\hline\hline
States & $J_z$ & Bases $(Y_{LL_z} \chi_{S S_z})$ \\
\hline
$h_c \ (^1 \! P_1)$ & $0$    & $ Y_{10}  \chi_{00} $ \\
        & $\pm1$ & $ Y_{1\pm1} \chi_{00} $ \\
$\chi_{c0} \ (^3 \! P_0)$ & $0$    & $ \frac{1}{\sqrt{3}} [Y_{11} \chi_{1-1} - Y_{10} \chi_{10} +Y_{1-1} \chi_{11}] $ \\
$\chi_{c1} \ (^3 \! P_1)$ & $0$    & $ \frac{1}{\sqrt{2}} [Y_{1-1} \chi_{11} - Y_{11} \chi_{1-1}]$ \\
        & $\pm1$ & $ \pm \frac{1}{\sqrt{2}} [Y_{10} \chi_{1\pm1} - Y_{1\pm 1} \chi_{10}]$ \\
$\chi_{c2} \ (^3 \! P_2)$ & $0$ & $ \frac{1}{\sqrt{6}} [Y_{11} \chi_{1-1} + 2 Y_{10} \chi_{10} + Y_{1-1} \chi_{11} ] $ \\
  & $\pm1$ & $ \frac{1}{\sqrt{2}} [Y_{1 \pm1} \chi_{10} + Y_{10} \chi_{1\pm1}] $ \\
  & $\pm2$ & $ Y_{1 \pm1} \chi_{1\pm1} $ \\
\hline\hline
\end{tabular}
\label{Tab_CG_vac}
\end{table}

\begin{figure}[t!]
    \centering
    \includegraphics[clip, width=1.0\columnwidth]{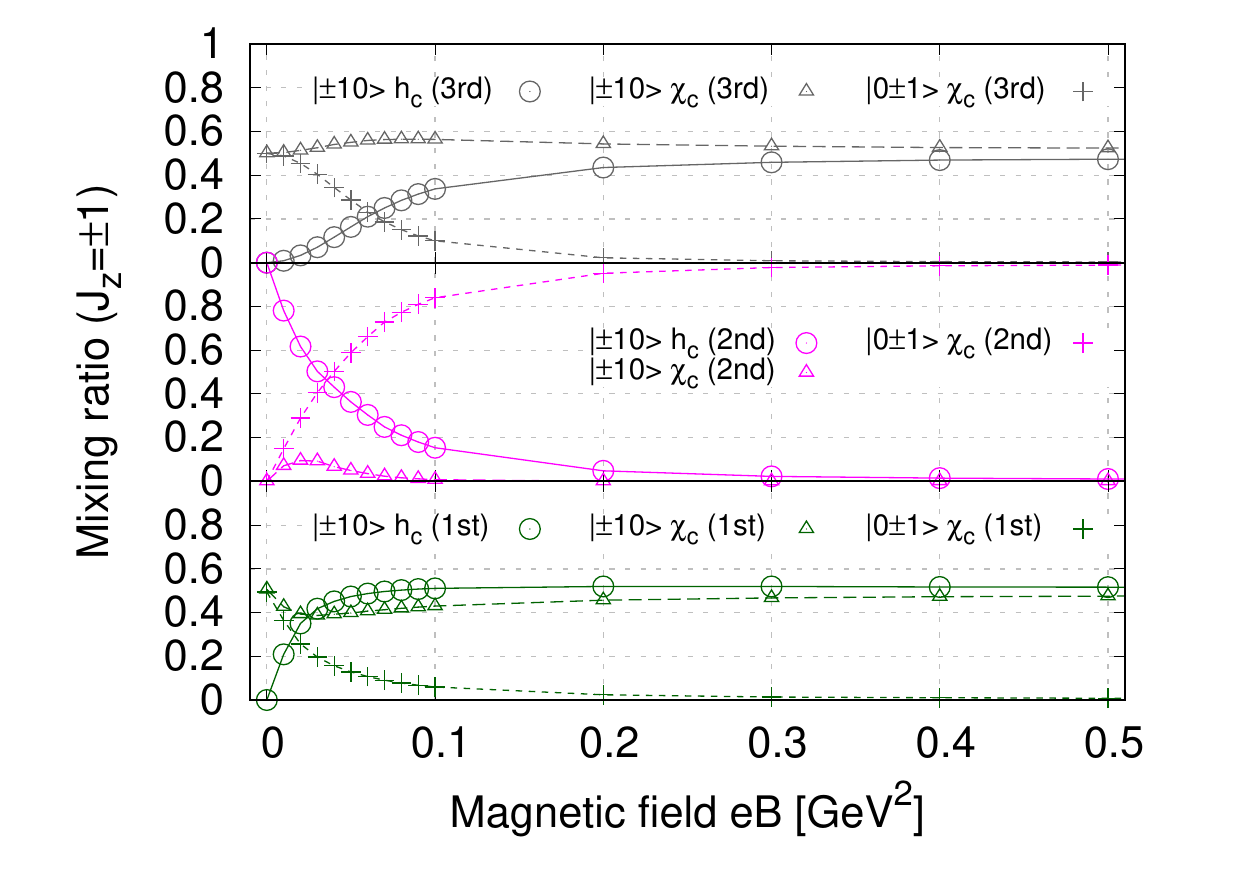}
    \includegraphics[clip, width=1.0\columnwidth]{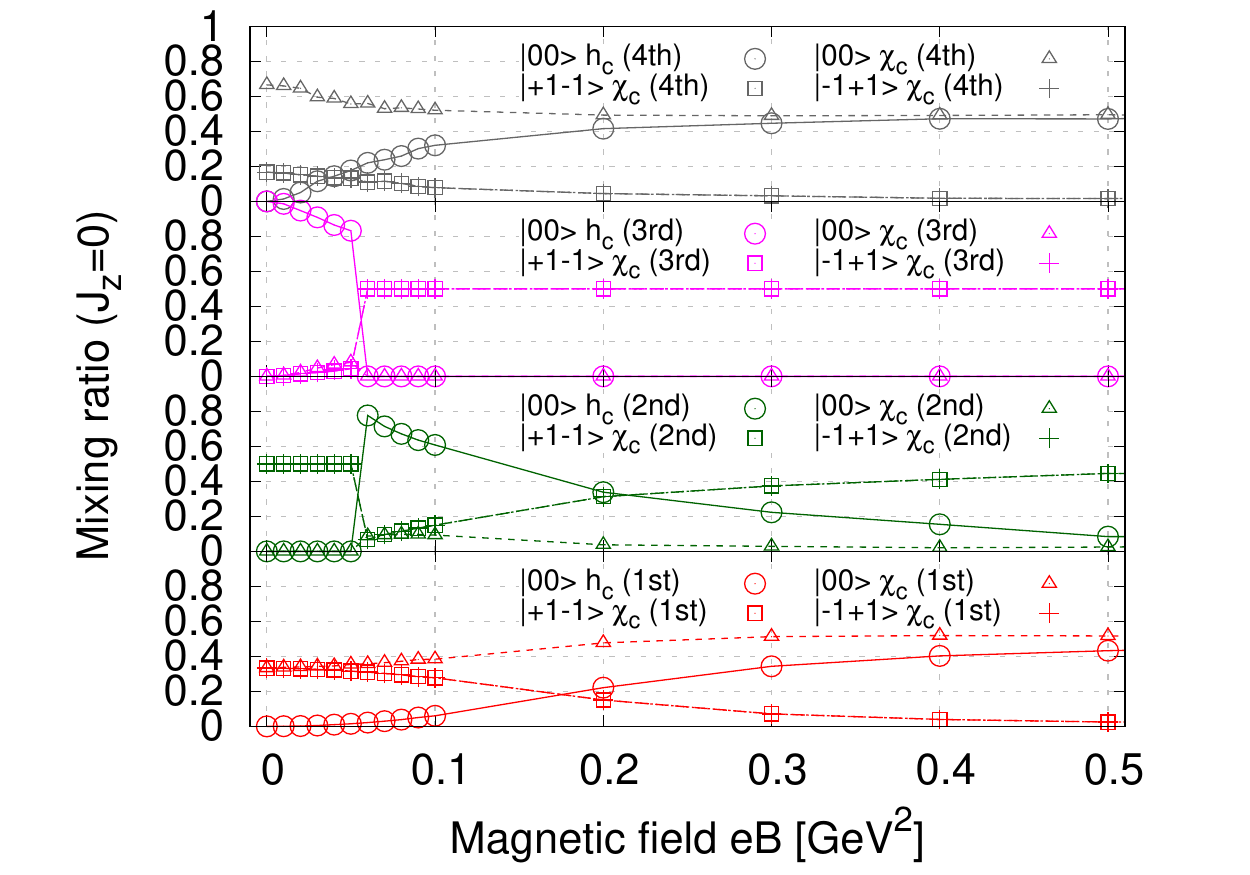}
    \caption{Mixing ratios among the $Y_{LL_z} \chi_{S S_z}$ basis for the P-wave charmonia in a magnetic field. Upper: $J_z=\pm1$. Lower: $J_z=0$.The legends stand for $\ket{L_zS_z} \chi_{SS_z} $ bases, where the spin components, ``$h_c$" and ``$\chi_c$", correspond to $\chi_{00}$ and $\chi_{1S_z}$, respectively.}
    \label{Fig_rate2}
\end{figure}

\begin{table}[t!]
\centering
\caption{Wave functions of P-wave charmonia {\it in the PB (strong-field) limit}, represented by $Y_{LL_z} \chi_{S S_z}$ basis.}
\begin{tabular}{cc}
\hline\hline
$J_z$ & Bases $(Y_{LL_z} \chi_{S S_z})$ \\
\hline
$0$  & $ \frac{1}{\sqrt{2}} [Y_{10}  \chi_{00} + Y_{10} \chi_{10}] $ \\
$0$  & $ \frac{1}{\sqrt{2}} [Y_{11} \chi_{1-1} + Y_{1-1} \chi_{11}] $ \\
$0$  & $ \frac{1}{\sqrt{2}} [Y_{11} \chi_{1-1} - Y_{1-1} \chi_{11}] $ \\
$0$  & $ \frac{1}{\sqrt{2}} [Y_{10}  \chi_{00} - Y_{10} \chi_{10}] $ \\
\hline
$\pm1$ & $ \frac{1}{\sqrt{2}} [Y_{1\pm1}  \chi_{00} +  Y_{1\pm 1} \chi_{10}] $ \\
$\pm1$ & $ Y_{10} \chi_{1\pm1}$  \\
$\pm1$ & $ \frac{1}{\sqrt{2}} [Y_{1\pm1}  \chi_{00} -  Y_{1\pm 1} \chi_{10}] $\\
\hline
$\pm2$ & $ Y_{1 \pm1} \chi_{1\pm1} $ \\
\hline\hline
\end{tabular}
\label{Tab_CG_PB}
\end{table}

\begin{figure}[t!]
    \centering
    \includegraphics[clip, width=1.0\columnwidth]{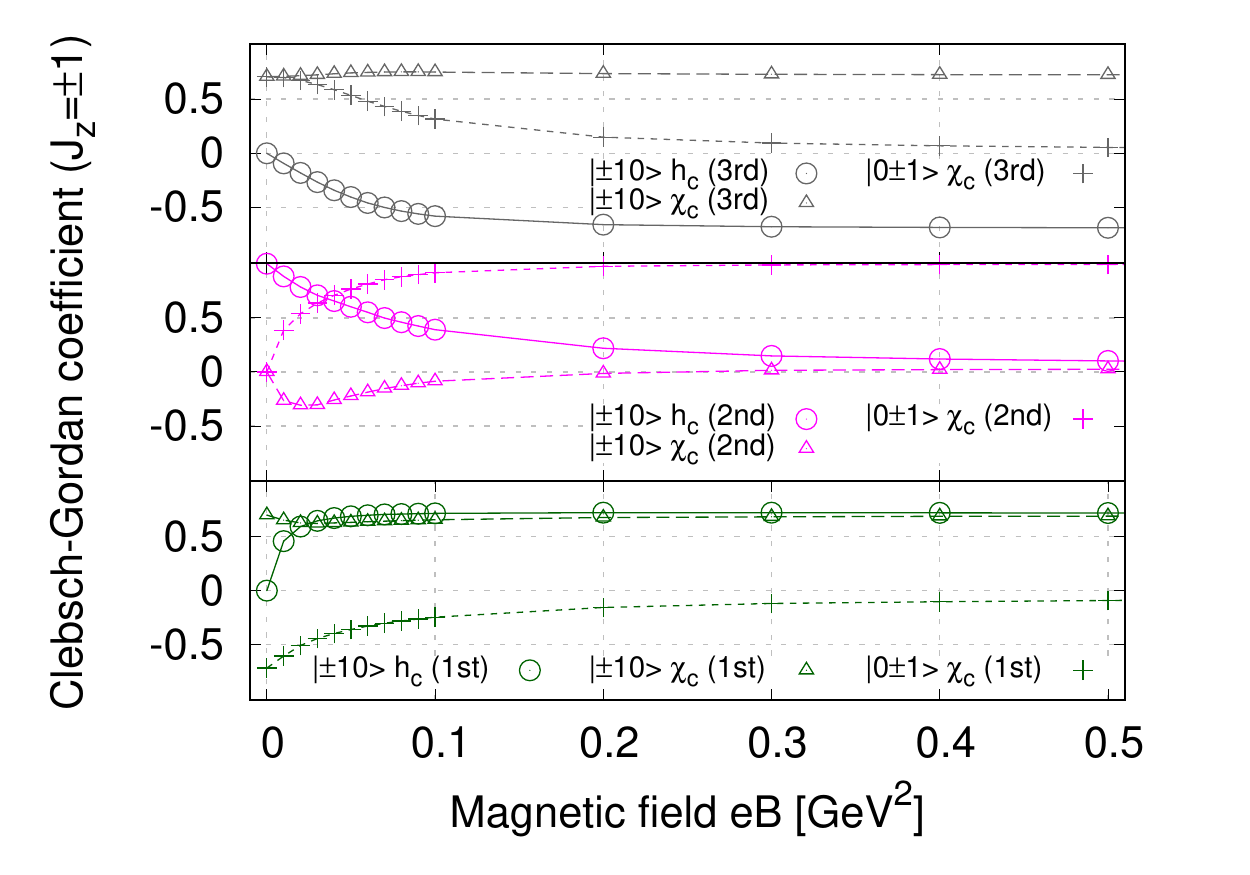}
    \includegraphics[clip, width=1.0\columnwidth]{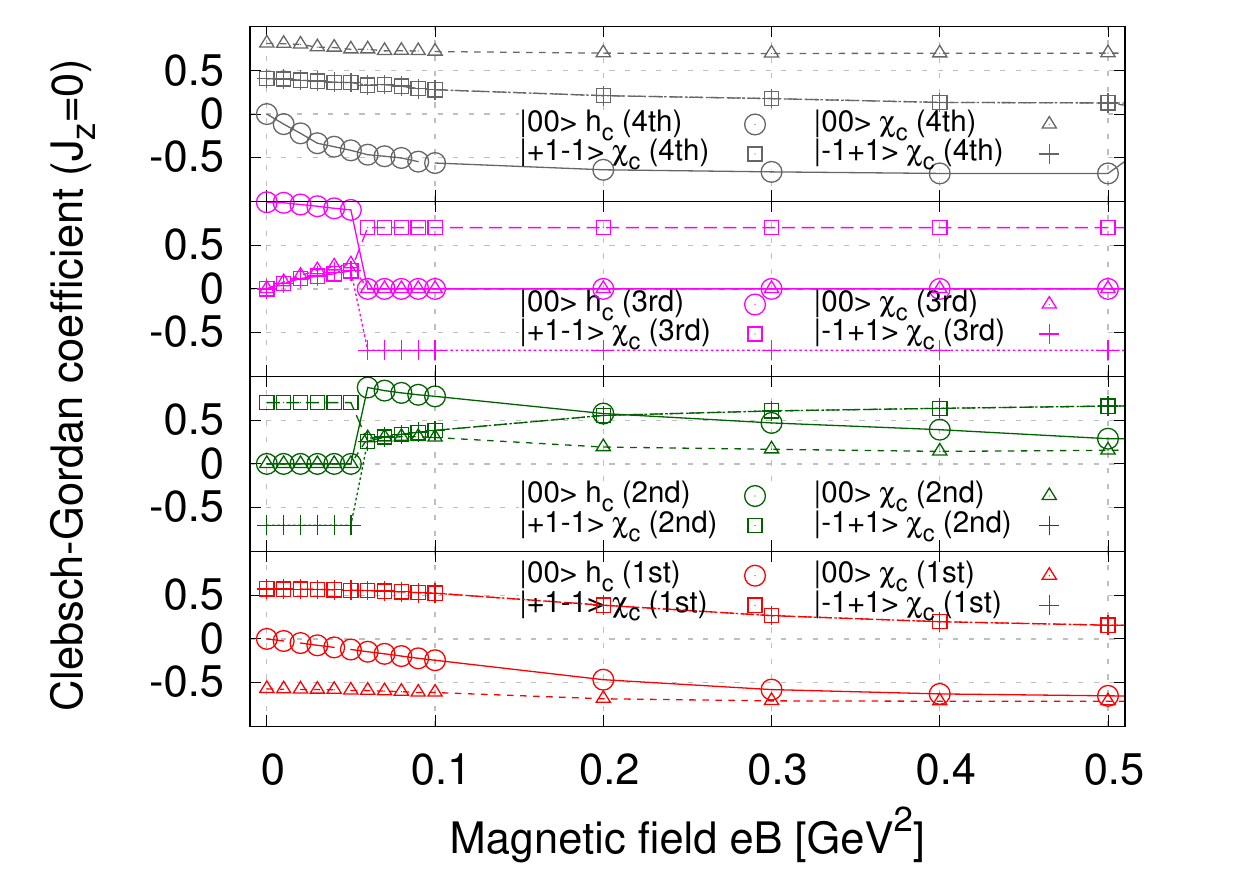}
    \caption{Clebsch-Gordan coefficients for the $Y_{LL_z} \chi_{S S_z}$ bases of P-wave charmonia in a magnetic field. Upper: $J_z=\pm1$. Lower: $J_z=0$.}
    \label{Fig_CGcoe}
\end{figure}

\section{Clebsch-Gordan coefficients in a magnetic field}

The Clebsch-Gordan (CG) coefficients for the wave functions of P-wave charmonia in vacuum are summarized in Table~\ref{Tab_CG_vac}, where we used $Y_{LL_z} \chi_{S S_z}$ basis.
On the other hand, the expected bases in the PB limit, where the mixing by the LS coupling can be neglected, are summarized in Table~\ref{Tab_CG_PB}.
Here, the states with $S_z=0$, $Y_{10} \chi_{00}$ and $Y_{10} \chi_{10}$ for $J_z=0$ and $Y_{1\pm1} \chi_{00}$ and $Y_{1\pm 1} \chi_{10}$ for $J_z=\pm1$, are mixed by the magnetic moments.
Furthermore, $Y_{11} \chi_{1-1}$ and $Y_{1-1} \chi_{11}$ for $J_z=0$ are mixed by the tensor coupling even in the PB limit.

The numerical results of the mixing ratios and CG coefficients for the $Y_{LL_z} \chi_{S S_z}$ basis in zero and nonzero magnetic fields are shown in Figs.~\ref{Fig_rate2} and \ref{Fig_CGcoe}, respectively.
From these figures, we see that the CG coefficients in vacuum on Table~\ref{Tab_CG_vac} are successfully reproduced.
In the strong magnetic field (PB) limit, the CG coefficients converge into the constant values as Table~\ref{Tab_CG_PB}.

\bibliography{HPBE}

\onecolumn
\section{Supplementary material for: ``Hadronic Paschen-Back effect"}
\subsection*{List of wave functions in a magnetic field}

\begin{figure*}[h!]
    \centering
    \includegraphics[clip, width=18cm]{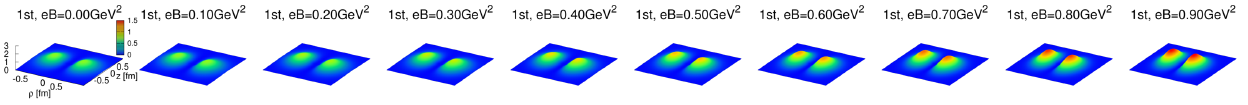}
    \vspace{-10pt}
    \caption{Probability densities of wave functions of P-wave charmonia with $J_z=\pm2$ in a magnetic field.}
    \label{Fig_sup_WFJz2}
    \vspace{-10pt}
\end{figure*}

\begin{figure*}[h!]
    \centering
    \includegraphics[clip, width=18cm]{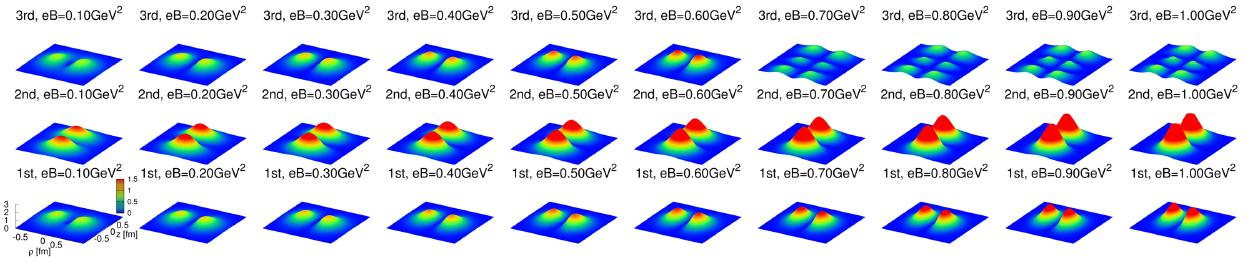}
    \includegraphics[clip, width=18cm]{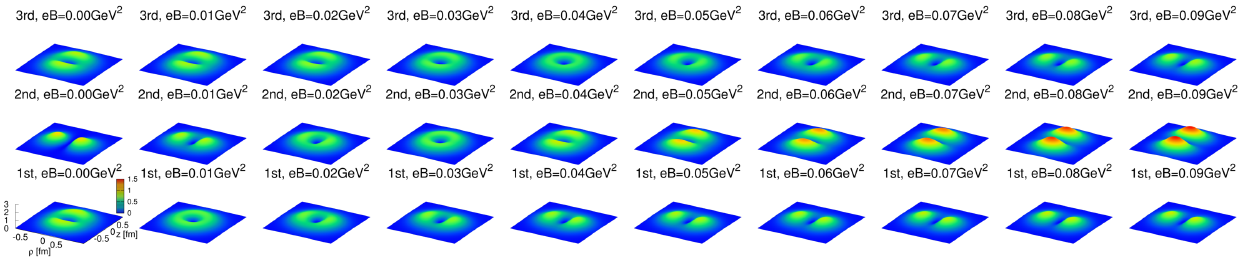}
    \vspace{-10pt}
    \caption{Probability densities of wave functions of P-wave charmonia with $J_z=\pm1$ in a magnetic field.}
    \label{Fig_sup_WFJz1}
    \vspace{-10pt}
\end{figure*}

\begin{figure*}[h!]
    \centering
    \includegraphics[clip, width=18cm]{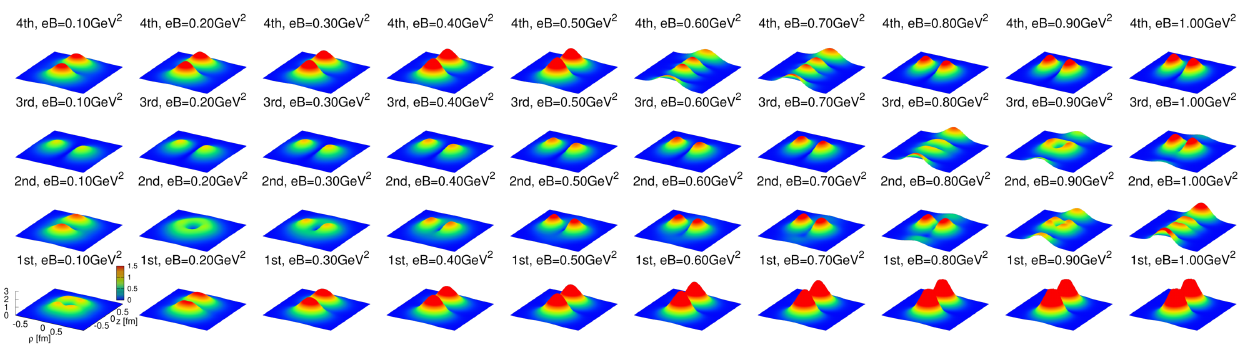}
    \includegraphics[clip, width=18cm]{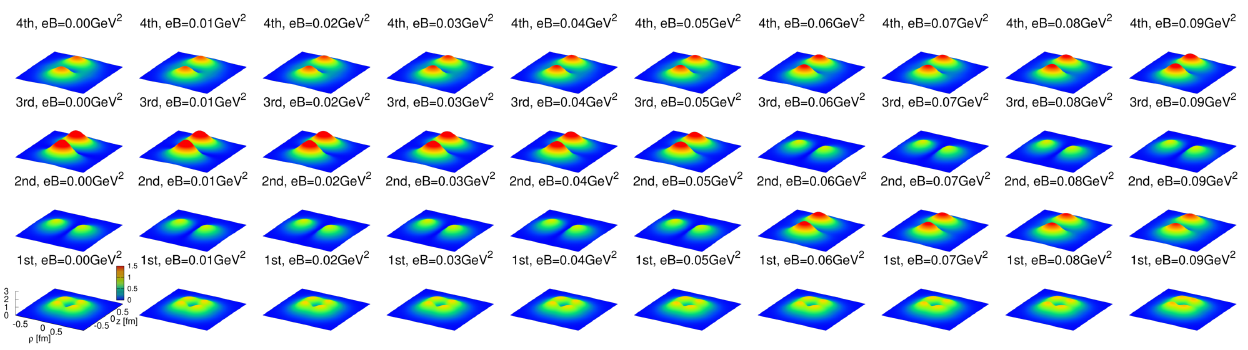}
    \vspace{-10pt}
    \caption{Probability densities of wave functions of P-wave charmonia with $J_z=0$ in a magnetic field.}
    \label{Fig_sup_WFJz0}
    \vspace{-10pt}
\end{figure*}

In the Figs.~\ref{Fig_sup_WFJz2}, \ref{Fig_sup_WFJz1}, and \ref{Fig_sup_WFJz0}, we show all the wave functions for $J_z=\pm 2, \pm 1, 0$ channels, respectively.
The corresponding mass spectra and mixing ratios are shown in the main text.

\end{document}